\documentclass[twocolumn,superscriptaddress,amsmath,amssymb,aps,pra]{revtex4}
\usepackage{amsmath,amsthm,amssymb}
\usepackage{latexsym}
\usepackage{amscd,graphicx}
\usepackage[titletoc]{appendix}
\usepackage{epsfig}
\usepackage{epstopdf}
\usepackage[normalem]{ulem}
\usepackage[dvipsnames]{xcolor}
\usepackage{booktabs}
\usepackage[colorlinks=true,linkcolor=red,citecolor=blue]{hyperref}

\newcommand{\norm}[1]{\| #1 \|}
\newcommand{\abs}[1]{| #1 |}

\begin{document}

\title{Exponentially Enhanced Tripartite Coupling in Quantum Nonlinear Magnonics}

\author{Xue-Chun Chen}
\affiliation{Department of Physics, Wenzhou University, Zhejiang 325035, China}

\author{Zi-Jie Wang}
\affiliation{Department of Physics, Wenzhou University, Zhejiang 325035, China}

\author{Sheng-Bo Zheng}
\affiliation{Department of Physics, Wenzhou University, Zhejiang 325035, China}

\author{Jiaojiao Chen}
\altaffiliation{jjchenphys@hotmail.com}
\affiliation{Department of Physics, Wenzhou University, Zhejiang 325035, China}

\author{Wei Xiong}
\altaffiliation{xiongweiphys@wzu.edu.cn}
\affiliation{Department of Physics, Wenzhou University, Zhejiang 325035, China}
\affiliation{International Quantum Academy, Shenzhen, 518048, China}

\date{\today }

\begin{abstract}
Strong and controllable tripartite interactions play a pivotal role in quantum information and nonlinear quantum optics, yet challenging to realize. In this work, we propose a hybrid system consisting of a nitrogen-vacancy (NV) center coupled to Kerr magnons (magnons with Kerr nonlinearity) in two yttrium-iron-garnet spheres. By adiabatically eliminating the ground state of the NV qutrit in the dispersive regime, an effective tripartite interaction among magnons and an NV qubit encoded in its excited states is obtained. In the strong driving limit, Kerr magnons can be linearized and give rise to degenerate parametric amplification for squeezing magnons. As a result, both the tripartite interaction and cooperativity are exponentially enhanced twice, which is about $\exp(\xi)$ times than schemes only involving  single-squeezing. Hence, our proposal is more experimentally feasible because modest squeezing parameter is sufficient. With this amplified tripartite coupling strength, the system dynamics are greatly accelerated, leading to fast generation of tripartite entanglement. In addition, noise-resilient perfect magnon blockade can be achieved, well predicted by both the analytical approach and numerical simulation with quantum master equation.  Our results suggest that the NV center represents a promising interface for engineering many-body interactions in quantum magnonics, offering a versatile platform for exploring fundamental quantum phenomena such as entanglement and correlations.
\end{abstract}

\maketitle

\section{introduction}

Magnons~\cite{rameshti2022cavity,lachance2019hybrid,yuan2022quantum,prabhakar2009spin,van1958spin}, the quanta of spin waves, are collective excitations of spins in ferrimagnetic materials such as yttrium iron garnet (YIG). Owing to their intrinsically low dissipation~\cite{huebl2013high,tabuchi2014hybridizing,zhang2014strongly,zhang2015cavity}, high controllability~\cite{zhang2017observation,zhang2019experimental}, and strong hybridization with other quantum platforms, including microwave cavities~\cite{huebl2013high,tabuchi2014hybridizing,zhang2014strongly,zhang2015cavity} and superconducting qubits~\cite{tabuchi2015coherent,xu2023quantum,xu2024macroscopic,weng2025quantum}, magnons have recently attracted significant interest in quantum information science~\cite{yuan2022quantum,li2020hybrid}. These developments have given rise to the fields of cavity magnonics~\cite{rameshti2022cavity} and quantum magnonics~\cite{yuan2022quantum}.

Over the past decade, cavity magnonics has been extensively explored, revealing strong coupling phenomena~\cite{huebl2013high,tabuchi2014hybridizing,zhang2014strongly,zhang2015cavity} and rich non-Hermitian physics~\cite{wang2020dissipative,xu2019cavity,harder2021coherent,wang2019nonreciprocity,harder2018level}. Toward quantum magnonics, one promising route is to couple magnons and superconducting qubits through a common cavity mode~\cite{tabuchi2015coherent,xu2023quantum,xu2024macroscopic}, where the cavity mediates effective indirect interactions. Alternatively, magnons can be interfaced with solid-state spin systems such as nitrogen-vacancy (NV) centers~\cite{neuman2020nanomagnonic,xiong2022strong,fukami2021opportunities,candido2021predicted,cornelissen2015long,ji2023kerr,skogvoll2021tunable,karanikolas2022magnon,hetenyi2022long} or magnetic skyrmions~\cite{pan2024magnon,li2022interaction}. Compared with superconducting qubits, solid-state spins offer substantially longer coherence times~\cite{balasubramanian2009ultralong,bar-gill2013solid,neumann2008multipartite}, a key resource for quantum information processing and computation~\cite{yao2012scalable,pfender2017nonvolatile,barry2020sensitivity,barson2017nanomechanical,nemoto2014photonic,ladd2010quantum,shi2015single,kucsko2013nanometre,li2025observation}.

To date, most studies in quantum magnonics have focused on pairwise interactions, analogous to the Jaynes–Cummings model describing a qubit coupled to a single-mode cavity~\cite{jaynes2005comparison}. Such couplings have enabled numerous breakthroughs and remain central to quantum science. With the rapid advancement of quantum technologies, however, there is growing interest in tripartite interactions that extend beyond pairwise coupling and could enable more advanced quantum functionalities~\cite{pan2025tripartite,hei2023enhanced,tang2024unveiling,cotrufo2017coherent,xu2019generation,wang2022unconventional,shao2023frequency,zhou2022synergistic}. Achieving strong and tunable tripartite interactions nevertheless remains a major challenge.

In this work, we propose a scheme to realize strong and tunable tripartite interactions in hybrid quantum magnonics, where a three-level nitrogen–vacancy (NV) center (i.e., a qutrit) couples to driven magnons in two spatially separated yttrium iron garnet (YIG) spheres. By driving magnons far off-resonant with the NV qutrit, the ground state of the qutrit can be adiabatically eliminated, thereby inducing an effective interaction among the magnons and an NV qubit encoded in the two excited states. To further enhance this interaction, we exploit the Kerr nonlinearity of magnons (i.e., Kerr magnons) in the YIG spheres, originating from magnetocrystalline anisotropy~\cite{zhang2019theory,wang2016magnon}. Such Kerr effects have been widely investigated in contexts ranging from nonlinear stability~\cite{wang2018bistability,nair2020nonlinear,shen2021long} and nonreciprocity~\cite{kong2019magnon,xiong2023highly} to quantum entanglement~\cite{zhang2019quantum,chen2023nonreciprocal,chen2024nonreciprocal,liu2025nonreciprocal}, quantum phase transitions~\cite{liu2023switchable,zhang2021parity}, strong spin–spin coupling~\cite{xiong2022strong,peng2025cavity,xiong2023optomechanical}, and ground-state cooling~\cite{fan2025hybrid}.  
Under strong driving, we linearize the Kerr magnons, leading to degenerate parametric amplification (DPA). Applying a squeezing transformation converts them into squeezed magnons, in which both the tripartite interaction strength and cooperativity are enhanced by $\exp(2\xi)$ due to simultaneous squeezing of the two modes. This joint squeezing yields an additional $\exp(\xi)$ enhancement compared with conventional single-mode squeezing, thereby relaxing experimental requirements: only modest squeezing suffices, whereas single-squeezing schemes typically demand much larger parameters.  
With this enhanced interaction, we show that system dynamics are significantly accelerated, enabling fast generation of tripartite entanglement. We further analyze the impact of dissipation on both entanglement dynamics and fidelity. In addition, we demonstrate that the enhanced interaction facilitates noise-resilient perfect magnon blockade, supported by analytical calculations and numerical simulations. Our results establish multilevel spin–magnon coupling as a powerful route toward quantum magnonics with tripartite interactions, opening opportunities for the exploration of exotic quantum phenomena.

\section{MODEL AND HAMILTONIAN}\label{s2}
\begin{figure}
	\includegraphics[scale=0.4]{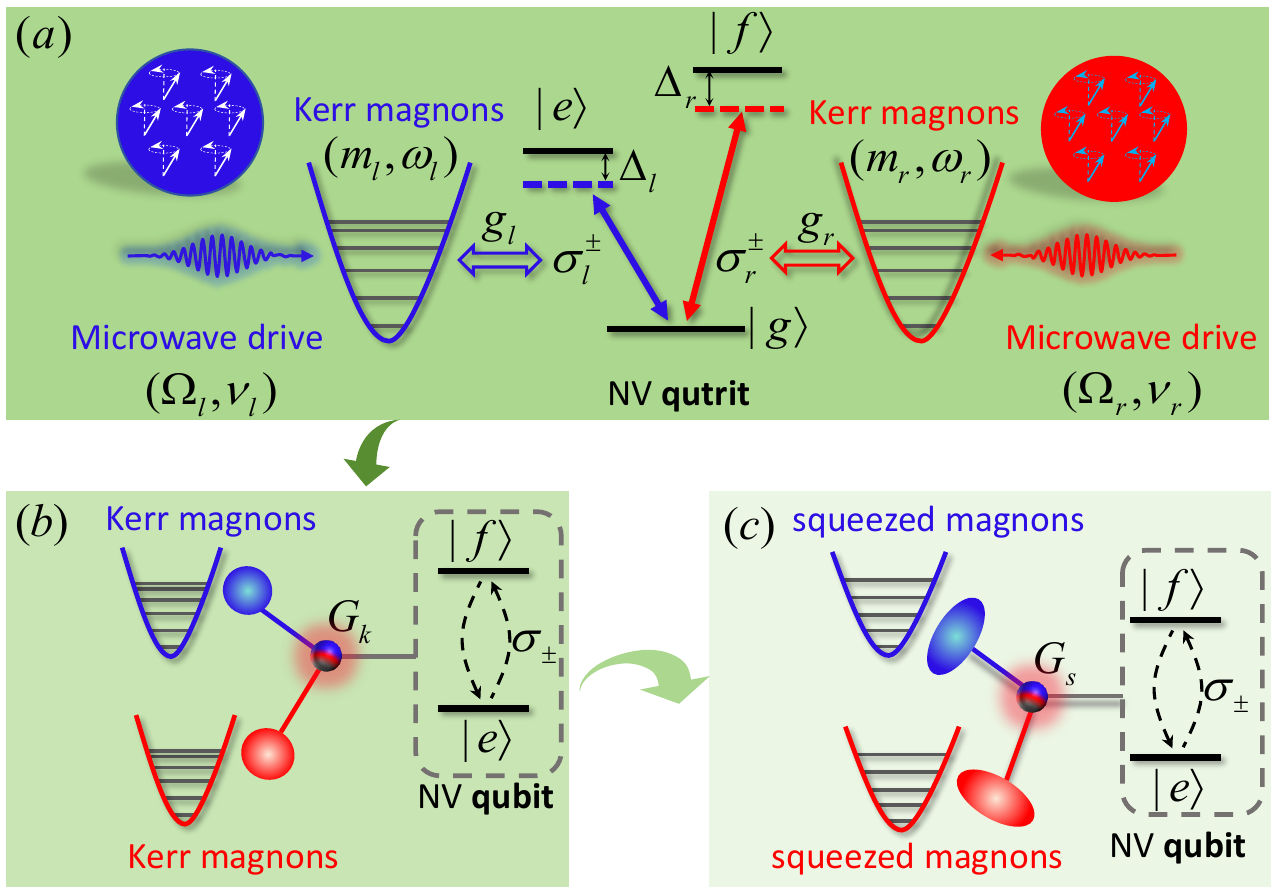}  
	\caption{(a) Schematic of the hybrid system consisting of an NV center coupled to Kerr magnons in two YIG spheres. magnons in each sphere are driven by a microwave field with frequency $\nu_p$ and amplitude $\Omega_p$ ($p=l,r$). The transition $|g\rangle\!\leftrightarrow\!|e\rangle$ ($|g\rangle\!\leftrightarrow\!|f\rangle$) couples to the magnons in the left (right) sphere.	(b) Effective tripartite interaction between the the magnons in two spheres and the NV qubit formed by two excited states of the NV qutrit, obtained by adiabatic elimination of the ground state $|g\rangle$ in the dispersive regime. (c) Exponentially enhanced tripartite interaction arising from magnon squeezing in both spheres, induced by the Kerr nonlinearity. } \label{fig1}
\end{figure}
We consider a hybrid quantum system composed of two driven YIG spheres and a NV center in diamond, as illustrated in Fig.\,\ref{fig1}(a). The NV center has three levels, denoted by $|g\rangle$, $|e\rangle$, and $|f\rangle$, acting as a $V$-type qutrit. Note that the levels of $|e\rangle$ and $|f\rangle$ are degenerate without an external bias magnetic field, separated from the state $|g\rangle$ by a zero-field splitting energy $D/2\pi \approx 2.87\,\mathrm{GHz}$~\cite{schirhagl2014nitrogen}. Each YIG sphere supports Kerr magnons, i.e., magnons with intrinsic Kerr nonlinearity, which can be effectively viewed as anharmonic oscillators-like. The Kerr nonlinearity originates from the magnetocrystalline anisotropy of the YIG material, and its strength can be tuned by adjusting both the direction and the amplitude of the applied static magnetic field~\cite{wang2016magnon,wang2018bistability}.  The total Hamiltonian of the system is written as (setting $\hbar=1$)
\begin{align}
	H_{\rm total}=H_{\rm nv}+H_{\rm km}+H_{\rm int}+H_{\rm drive},\label{equ1}
\end{align}
where
\begin{align}
	H_{\rm nv}=\omega_g|g\rangle\langle g|+\omega_e|e\rangle\langle e|+\omega_f|f\rangle\langle f|
\end{align}
is the free Hamiltonian of the NV qutrit ($\omega_g$ is the energy of the state $|g\rangle$, $\omega_e$ is the energy of the state $|e\rangle$, and $\omega_f$ is the energy of the state $|f\rangle$). The second term in Eq.\,(\ref{equ1}),
\begin{align}
	H_{\rm km}=\sum_{p=l,r}\omega_p m_p^\dagger m_p-\frac{K_p}{2}(m_p^\dagger m_p)^2,
\end{align} 
is the Hamiltonian of the Kerr magnons, with $\omega_p$ being the angular frequency of the magnons, $K_p$ the strength of the Kerr nonlinearity, and $m_p$ ($m_p^\dagger$) the annihilation (creation) operator of the magnons. The third term $H_{\rm int}$ in Eq.\,(\ref{equ1}) describes the interaction between the Kerr magnons in two spheres and the NV qutrit. Specifically, the magnons in the left YIG sphere is coupled to the transition between $|g\rangle$ and $|e\rangle$ with the coupling strength $g_l$, and the magnons in the right YIG sphere is coupled to the transition between $|g\rangle$ and $|f\rangle$ with the coupling strength $g_r$. Within the rotating-wave approximation (RWA), the Hamiltonian $H_{\rm int}$ takes the form
\begin{align}
	H_{\rm int}=\sum_{p=l,r}g_p (m_p\sigma_p^++m_p^\dagger\sigma_p^-),
\end{align}
where $\sigma_l^-=|g\rangle \langle e|$ ($\sigma_l^+=|e\rangle \langle g|$) is the lowering (rising) operator for the transition $|g\rangle \to\langle e|$ ($|e\rangle \to\langle g|$), and $\sigma_r^-=|g\rangle \langle f|$ ($\sigma_r^+=|f\rangle \langle g|$) is the lowering (rising) operator for the transition $|g\rangle \to\langle f|$ ($|f\rangle \to\langle g|$). The last term in Eq.\,(\ref{equ1}),
\begin{align}
	H_{\rm drive}=\sum_{p=l,r}\Omega_p \left[m_p\exp(i\nu_pt)+m_p^\dagger\exp(-i\nu_pt)\right] 
\end{align}
characterizes the interaction between the classical microwave field and the magnons with RWA, where $\Omega_p$ is the amplitude and $\nu_p$ is the frequency.

We first rotate both the magnons in the left (right) sphere and the NV qutrit involving the transition $|g\rangle\leftrightarrow |e\rangle$ ($|g\rangle\leftrightarrow|f\rangle$) into the frame with respect to the frequency $\nu_l$ ($\nu_r$) of the left (right) driving field,  i.e.,  $m_l\rightarrow m_l e^{-i\omega_l t}$, $m_r\rightarrow m_r e^{-i\omega_r t}$, $\sigma_l^-\rightarrow \sigma_l^- e^{-i\omega_l t}$, and $\sigma_r^-\rightarrow \sigma_r^- e^{-i\omega_r t}$. In the rotating frame, the Hamiltonian of the system in Eq.\,(\ref{equ1}) reduces to
\begin{align}
	H_{\rm total}^\prime=H_{\rm nv}^\prime+H_{\rm km}^\prime+H_{\rm int}^\prime+H_{\rm drive}^\prime,\label{equ6}
\end{align}
with
\begin{align}
	H_{\rm nv}^\prime=&\omega_g|g\rangle\langle g|+\Delta_e|e\rangle\langle e|+\Delta_f|f\rangle\langle f|,\\
	H_{\rm km}^\prime=&\sum_{p=l,r}\left[\delta_p m_p^\dagger m_p-\frac{K_p}{2}(m_p^\dagger m_p)^2\right],\\
	H_{\rm drive}^\prime=&\sum_{p=l,r}\Omega_p \left(m_p+m_p^\dagger\right),~{\rm and}~H_{\rm int}^\prime=H_{\rm int},
\end{align}
where $\Delta_e=\omega_e-\nu_l$, $\Delta_f=\omega_e-\nu_r$, and $\delta_p=\omega_p-\nu_p$ are frequency detunings.

Since the Kerr nonlinearity of magnons is included, we next perform a standard linearization under the assumption of strong coherent driving fields ($|\Omega_p|\gg1$). In this regime, the magnon operators can be displaced as $m_p \rightarrow \langle m_p\rangle + m_p$, where $\langle m_p\rangle$ is the classical steady-state amplitude and $m_p$ here denotes the quantum fluctuation. After linearization, the Kerr nonlinearity gives rise to a frequency shift to the magnons and induces DPA, that is, the Hamiltonian $H_{\rm km}^\prime$ in Eq.\,(\ref{equ6}) becomes~\cite{xiong2022strong}
\begin{align}
	H_k^\prime
	= \sum_{p=l,r} \left[\omega_{k,p} m_p^\dagger m_p
	- \frac{1}{2}\mathcal{K}_p\left(m_p^{\dagger 2} + m_p^2\right)\right],\label{equ10}
\end{align}
where $\omega_{k,p}=\delta_p + \delta_{k,p}$ is the modified frequency of the Kerr magnons, $\delta_{k,p} = -2K_p|\langle m_p\rangle|^2$ represents the nonlinear frequency shift, and $\mathcal{K}_p = K_p \langle m_p\rangle^2$ is the tunable coefficient of DPA by control of the amplitude of the driving field.

 \begin{figure}
	\includegraphics[scale=0.26]{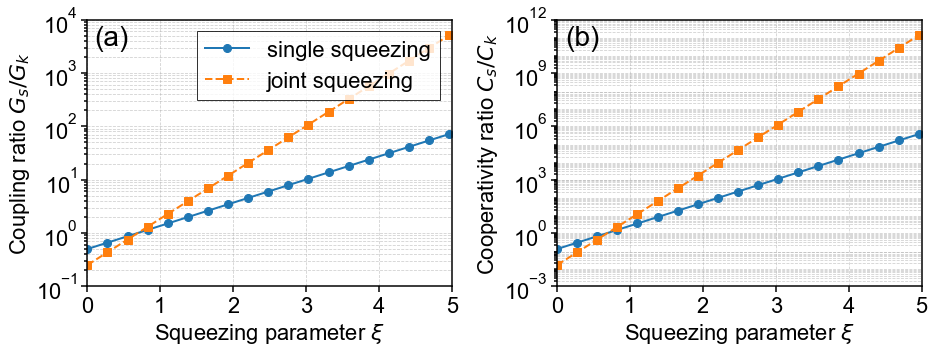}  
	\caption{(a) Coupling ratio $G_s/G_k$ and (b) cooperativity ratio $C_s/C_k$ versus the squeezing parameter $\xi$. The orange curve denotes the joint squeezing and the blue curve denotes the single squeezing.}\label{figure2}
\end{figure}
\section{Exponentially enhanced magnon-NV-magnon tripartite interaction}\label{s3}

Below we focus on the case that magnons in two spheres are far-off resonant with the NV qutrit\,[see Fig.\,\ref{fig1}(a)], i.e.,
\begin{align}
	g_l\ll& \abs{\Delta_l=\Delta_e-\omega_g-\omega_{k,l}}, \\
	g_r\ll& \abs{\Delta_r=\Delta_f-\omega_g-\omega_{k,r}}.
\end{align}
In this regime, the ground state $|g\rangle$ is only virtually populated, thereby it can be adiabatically eliminated to introduce a tripartite interaction among Kerr magnons and the NV qubit formed by the states $|e\rangle$ and $|f\rangle$\,[see Fig.\,\ref{fig1}(b)]. To show this process, we first rewrite the Hamiltonian in Eq.\,(\ref{equ6}) in the strong driving limit as
\begin{align}
	\mathcal{H}_k=&\mathcal{H}_0+\mathcal{V}-\sum_{p=l,r}\frac{\mathcal{K}_p}{2}(m_p^{\dagger2}+ m_p^2),\label{equ15}
\end{align}
with
\begin{align}
	\mathcal{H}_0=&H_{\rm nv}^\prime+\omega_{k,l}m_l^\dagger m_l+\omega_{k,r}m_r^\dagger m_r,\\ 
	\mathcal{V}=&\sum_{p=l,r}g_p (m_p\sigma_p^++m_p^\dagger\sigma_p^-).
\end{align}
Then we apply a Schrieffer-Wolff transformation to the Hamiltonian $\mathcal{H}_k$ in Eq.\,(\ref{equ15}). The unitary operator takes the form of $\mathcal{U}=e^{-\mathcal{S}}$, with
\begin{align}
	\mathcal{S} = \sum_{p=l,r} \left( \frac{g_p}{\Delta_p} m_p^\dagger \sigma_p^- - \frac{g_p}{\Delta_p} m_p \sigma_p^+ \right)
\end{align}
satisfying
\begin{align}
	[\mathcal{H}_0,\mathcal{S}]+\mathcal{V}=0.
\end{align}
Up to the second order, an effective tripartite interaction among Kerr magnons in two spheres and the NV qubit is obtained [see Fig.\,\ref{fig1}(b)], as
\begin{align}\label{equ18}
	H_{\rm eff}=&\frac{1}{2}\omega_q\sigma_z+H_k^\prime+G_k(m_l m_r^\dagger\sigma_-+m_l^\dagger m_r\sigma_+),
\end{align}
where $\omega_q=\Delta_f-\Delta_e+{g_r^2}/{\Delta_r}-{g_l^2}/{\Delta_l}\approx \Delta_f-\Delta_e$ is the effective frequency of the NV qubit, $\sigma_z=|f\rangle\langle f|-|e\rangle\langle e|$ is the Pauli-$z$ operator of the NV qubit, $\sigma_+=|f\rangle\langle e|$ ($\sigma_-=|e\rangle\langle f|$) is the rising (lowering) operator of the NV qubit, and the effective coupling strength is $G_k=(g_lg_r/2)(1/\Delta_l+1/\Delta_r)$.

Since DPA is included in Eq.\,(\ref{equ18}), the effective system can be operated in the squeezed-magnon representation. This can be achieved by applying a Bogoliubov transform
\begin{align}
	m_p=&m_{s,p} \cosh \xi_p+m_{s,p}^{\dagger} \sinh \xi_p,\\
	m_p^\dagger=&m_{s,p}^\dagger \cosh \xi_p+m_{s,p} \sinh \xi_p,
\end{align}
to Eq.\,(\ref{equ18}). When the squeezing parameter is set as
\begin{align}
	\xi_p=\frac{1}{4} \ln\frac{\omega_{k,p}+\mathcal{K}_p}{\omega_{k,p}-\mathcal{K}_p},
\end{align}
the Hamiltonian $H_k^\prime$ in Eq.\,(\ref{equ18}) is diagonalized as
\begin{align}
	H_k^\prime\rightarrow H_{s,p}=\sum_{p=l,r}\omega_{s,p} m_{s,p}^\dagger m_{s,p},
\end{align}
where $\omega_{s,p}=\sqrt{\omega_{k,p}^2-\mathcal{K}_p^2}$ is the frequency of the squeezed-magnon. In terms of the operators $m_{s,p}$ and $m_{s,p}^\dagger$, the effective Hamiltonian $H_{\rm eff}$ in Eq.\,(\ref{equ18}) is transformed to~[see Fig.\,\ref{fig1}(c)]
\begin{align}
	H_{s,\rm eff}=&\frac{1}{2}\omega_q\sigma_z+\omega_{s,l}m_{s,l}^\dagger m_{s,l}+\omega_{s,r}m_{s,r}^\dagger m_{s,r}\notag\\
	&+G_s(m_{s,l}m_{s,r}^\dagger\sigma_-+m_{s,l}^\dagger m_{s,r}\sigma_+),\label{equ12}
\end{align}
where
\begin{align}\label{equ24}
	\frac{G_s}{G_k}=\frac{1}{4}e^{2\xi}
\end{align}
with $\xi_l=\xi_r=\xi$ and $\omega_{s,l}+\omega_q\approx\omega_{s,r}$. This means that the tripartite interaction among magnons and the NV qubit is exponentially enhanced by the joint magnon squeezing, as shown in Fig.\,\ref{figure2}(a). Due to this enhancement, the corresponding cooperativity, defined as $C_s=G_s^3/(\kappa_l\kappa_r\gamma_q)$, is also exponentially enhanced\,[see Fig.\,\ref{figure2}(b)], i.e.,
\begin{align}\label{equ25}
	\frac{C_s}{C_k}=(\frac{G_s}{G_k})^3=\frac{1}{64}e^{6\xi},
\end{align}
where $C_k=G_k^3/(\kappa_l\kappa_r\gamma_q)$ is the cooperativity without squeezing. 

 \begin{figure}
	\includegraphics[scale=0.26]{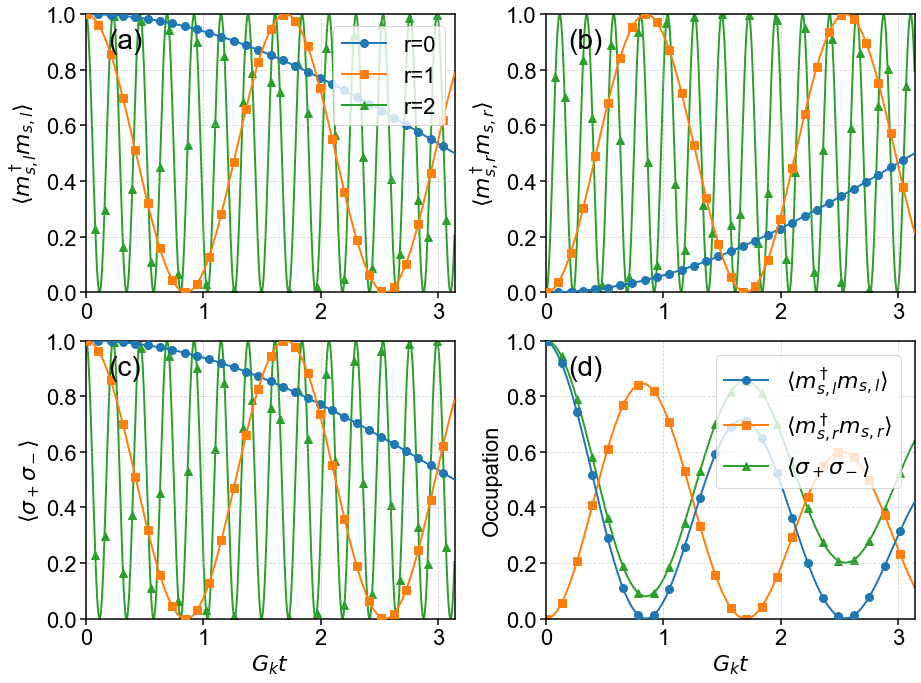}  
	\caption{The ideal occupation of (a) the left squeezed-magnon $\langle m_{s,l}^\dag m_{s,l}\rangle$, (b) the right squeezed-magnon $\langle m_{s,r}^\dag m_{s,r}\rangle$, and (c) the NV qubit $\langle \sigma_+\sigma_-\rangle$ as a function of the dimensionless evolution time $G_k t$. (d) The nonideal occupation of the left squeezed-magnon ($\langle m_{s,l}^\dag m_{s,l}\rangle$), the right squeezed-magnon ($\langle m_{s,r}^\dag m_{s,r}\rangle$), and the NV qubit ($\langle \sigma_+\sigma_-\rangle$) as a function of the dimensionless evolution time $G_k t$, where $\kappa_l/2\pi=\kappa_r/2\pi=100$ kHz, $\gamma_q/2\pi=1$ kHz and $\xi=2$. In (a)-(d),  the resonant condition $\omega_{s,l}+\omega_q=\omega_{s,r}$ is chosen, the system is initially prepared in the state $|\psi\rangle = |1_l,0_r,f_q\rangle_l$, and $G_k=0.5$\,MHz.} \label{figure3}
\end{figure}
From Eqs.\,(\ref{equ24}) and (\ref{equ25}), one can clearly see the advantage of the joint squeezing scheme in amplification of tripartite interaction. In particular, the joint squeezing approach significantly reduces the requirement for large squeezing parameters that are typically needed in single squeezing amplification but remain difficult to achieve experimentally. This can be clearly observed in Fig.\,\ref{figure2}. For example, squeezing parameters as large as $\xi=3\sim5$ are often assumed in single squeezing amplification schemes, the joint squeezing scheme can reach comparable performance with considerably smaller parameters, typically $\xi=1.5\sim2.5$. In other words, with the same squeezing parameter, the enhanced coupling strength and cooperativity are much larger with the joint squeezing approach than that with a single squeezing\,[see Figs.\,\ref{figure2}(a) and \ref{figure2}(b)]. 

To further demonstrate the pronounced enhancement of the tripartite coupling strength, Fig.\,\ref{figure3} presents the system dynamics initialized in the state $|\psi\rangle = |1_s\rangle_l |0_s\rangle_r |f\rangle_q$, where $|1_s\rangle_l$ denotes a single squeezed magnon in the left sphere, $|0_s\rangle_r$ denotes the vacuum state of the right squeezed magnon, and $|f\rangle_q$ represents the NV qubit in the state $|f\rangle$. As shown in Figs.\,\ref{figure3}(a)-\ref{figure3}(c), the enhanced tripartite coupling gives rise to faster Rabi oscillations compared to the unsqueezed case ($\xi=0$). Moreover, these results clearly demonstrate that the creation of a squeezed-magnon in the right sphere is necessarily accompanied by the annihilation of a squeezed-magnon in the left sphere, together with the transition of the NV qubit from $|f\rangle$ to $|e\rangle$. The dissipative dynamics in Fig.\,\ref{figure3}(d), governed by the Lindblad master equation $\dot{\rho}=i[\rho,H_{s,\rm eff}]+\kappa_l D[m_{s,l}]+\kappa_r D[m_{s,r}]+\gamma_qD[\sigma_-]$, further confirm that strong tripartite coupling can be achieved with relatively small squeezing parameters, e.g., $\xi=1$. The parameters $\kappa_l,\,\kappa_r$, and $\gamma_q$ represents the decay rates of the left squeezed-magnon, the right squeezed-magnon, and the NV qubit. $D[o]=2o\rho o^\dagger -\rho o^\dagger o-o^\dagger o \rho$ is the Lindblad superoperator.

\section{Tripartite entanglement}\label{s4}

\begin{figure}
	\includegraphics[scale=0.26]{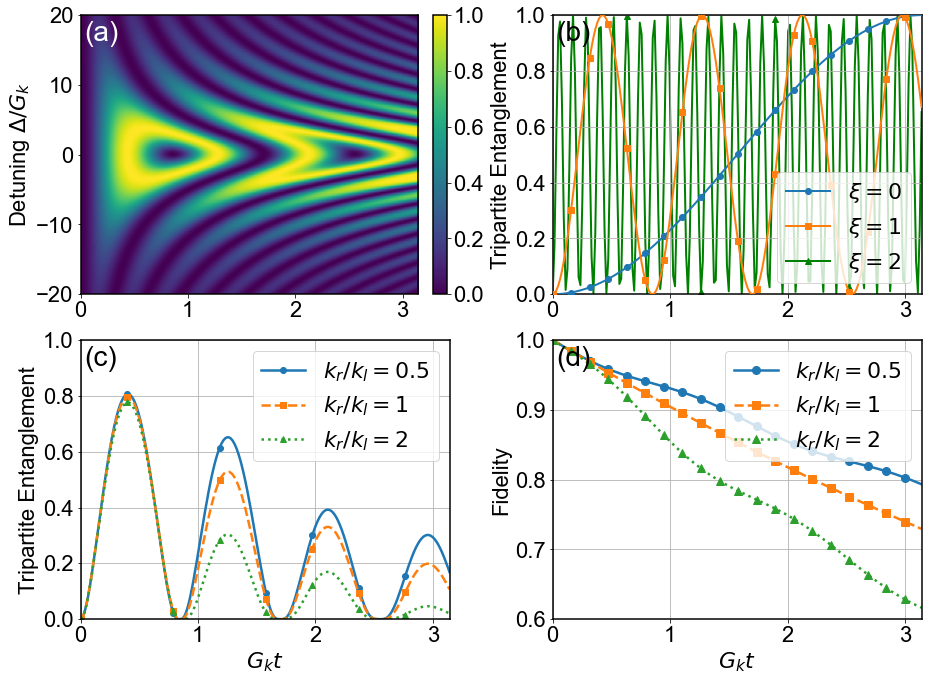}  
	\caption{(a) Tripartite entanglement versus time for different detuning quantified by the residual tangle $\tau_{A|B|C}$. (b) Tripartite entanglement versus time for different squeezing parameters $r$ by the residual tangle $\tau_{A|B|C}$. (c) Tripartite entanglement versus time for different decay rate of the system quantified by the residual tangle $\tau_{A|B|C}$ (d) Tripartite entanglement versus time quantified by the residual tangle $\tau_{A|B|C}$ and the minimum residual contangle $E_{\tau}^{i|j|k}$. Other parameters are the same as those in Fig.~\ref{figure2}.}\label{figure4}
\end{figure}

With Eq.\,(\ref{equ12}) at hand, tripartite entanglement among the left magnons, right magnons, and the NV qubit can be generated. To quantify this entanglement, we employ the minimum residual contangle, defined as
\begin{align}
	E_{\tau}^{A|B|C} \equiv \min_{A|B|C} \left[ E_{\tau}^{A|BC} - E_{\tau}^{A|B} - E_{\tau}^{A|C} \right],
\end{align}
where $(A,B,C)$ runs over all permutations of the tripartite system~\cite{adesso2006entanglement}. Here, $E_{\tau}^{A|BC} \equiv \log^2 \norm{\rho^p}$, with $\norm{\cdot}$ denoting the trace norm and the superscript '$p$' representing the partial transpose~\cite{vidal2002computable}. Similarly, $E_{\tau}^{A|B} = \log^2 \norm{ \rho_{AB}^p}$ and $E_{\tau}^{A|C} = \log^2 \norm{\rho_{AC}^p}$. To study the entanglement dynamics, the system is initially prepared in the state $|1_l,0_r,f_q\rangle$. Figure\,\ref{figure4}(a) shows the effect of the frequency detuning, $\Delta = \omega_{s,r} - \omega_{s,l} - \omega_q$, on the tripartite entanglement for a relatively small squeezing parameter $\xi=1$. It is evident that the maximum entanglement occurs at resonance (i.e., $\Delta=0$) and remains robust against detuning variations. Fixing $\Delta=0$, Fig.\,\ref{figure4}(b) illustrates the effect of the squeezing parameter on the entanglement dynamics. As $\xi$ increases from 0 to 2, the oscillation accelerates exponentially due to the exponential enhancement of the coupling strength. Figure\,\ref{figure4}(c) demonstrates that the tripartite entanglement exhibits a damped oscillatory behavior in the presence of dissipation. Larger ratios of $\kappa_r/\kappa_l$ lead to faster damping. In Fig.\,\ref{figure4}(d), we further investigate the effect of dissipation on the fidelity of the tripartite entangled state, defined as $\mathcal{F} = \langle {\rm GHZ}|\rho|{\rm GHZ}\rangle$, with $|{\rm GHZ}\rangle = (|1_l,0_r,f_q\rangle + |0_l,1_r,e_q\rangle)/\sqrt{2}$. The fidelity is nearly unaffected by $\kappa_r/\kappa_l$ at early times, but its degradation becomes significant as time evolves. Moreover, a larger $\kappa_r/\kappa_l$ results in a more rapid decay of entanglement. 

Note that the tripartite entanglement can also be characterized by the residual tangle $\tau_{A|B|C}$, defined as
\begin{align}
	\tau_{A|B|C}=4\abs{d_1-2d_2+4d_3},\label{equ27}
\end{align}
where $d_1$, $d_2$, and $d_3$ are given by Eq.~(21) in Ref.~\cite{coffman2000distributed}. Specifically, for the initial state $|1_l,0_r,f_q\rangle$ in our effective model, the time-evolved state can be obtained by solving the Schr$\ddot{o}$dinger equation. At time $t$, the state is
\begin{align}
	|\psi_t\rangle = &[\cos\left(\Omega t\right)
	- i\frac{\Delta}{2\Omega}\sin\left(\Omega t\right)]|1_l,0_r,f_q\rangle\notag\\
	& - i\frac{G_s}{\Omega}\sin\left(\Omega t\right)
	|0_l,1_r,e_q\rangle,
\end{align}
with $\Omega=\sqrt{G_s^2+\Delta^2/4}$. The corresponding density matrix is $\rho=|\psi_t\rangle\langle \psi_t|$, which determines the values of $d_1$, $d_2$, and $d_3$. Consequently, the residual tangle can be written as
\begin{align}
	\tau_{A|B|C}=\frac{4G_s^2}{\Omega^2}[\cos^2(\Omega t)+\frac{\Delta^2}{4\Omega^2}\sin^2(\Omega t)]\sin^2(\Omega t).
\end{align}
This expression reproduces the results shown in Figs.\,\ref{figure4}(a) and \ref{figure4}(b). Furthermore, the dissipative dynamics in Figs.\,\ref{figure4}(c) and \ref{figure4}(d) can be approximately described using the conditional Hamiltonian
\begin{align}
	H_{\rm con}=H_{s,\rm eff}-i\sum_{p=l,r}\frac{\kappa_{s,p}}{2}m_{s,p}^\dagger m_{s,p} -i\frac{\gamma_q}{2}\sigma_+\sigma_-.
\end{align}
These results indicate that the minimum residual contangle is equivalent to the residual tangle in quantifying tripartite entanglement.
 
\section{Magnon blockade based on quantum interference}\label{s5}

In addition to tripartite entanglement, the system described by Eq.\,(\ref{equ18}) can also be employed to explore magnon blockade. To this end, a weak probe field with amplitude $\Omega_p$ is applied to the magnons in the right sphere. Thus, the Hamiltonian in Eq.\,(\ref{equ18}) is modified as
\begin{align} \label{eq15}
	H_{\rm mb}&=H_{s,\rm eff}+\Omega_p(m_{s,r}^\dagger+m_{s,r}).
\end{align}
Since the field is weak ($\Omega_p \ll \kappa_{s,r}$), the magnon number in the right sphere is small, thus we can reasonably truncate the magnon number up to two. At time $t$, the state of the system can be written as
\begin{align} 
	\left|\Psi_t\right\rangle=&C_{00e} \left|0_l,0_r,e_q\right\rangle +C_{01e} \left|0_l,1_r,e_q\right\rangle+C_{02e} \left|0_l,2_r,e_q\right\rangle\notag\\
	&+C_{10f} \left|1_l,0_r,f_q\right\rangle +C_{11f} \left|1_l,1_r,f_q\right\rangle,
\end{align}
where $C_{ijk}$ with $i,j=0,\,1,\,2$ and $k=e,\,f$ are the probability amplitudes. By substituting the state $\left|\Psi_t\right\rangle$ into the Schrödinger equation, the following equations of motion for the probability amplitudes can be obtained,
\begin{align} \label{eq17}
	i\dot C_{00e}=&\Omega_p C_{01e}, \notag \\
	i\dot C_{01e}=&\tilde{\omega}_{s,l}C_{01e}+G_s C_{10f}+\Omega_p C_{00e}+\sqrt{2}\Omega_p C_{02e}, \notag\\
	i\dot C_{02e}=&2\tilde{\omega}_{s,l}C_{01e}+\sqrt{2}G_sC_{11f}+\sqrt{2}\Omega_p C_{01e}, \\
	i\dot C_{10f}=&\tilde{\omega}_q+\tilde{\omega}_{s,r}C_{10f}+G_sC_{01e}+\Omega_p C_{11f}, \notag\\
	i\dot C_{11f}=&(\tilde{\omega}_{s,l}+\tilde{\omega}_q+\tilde{\omega}_{s,r})C_{11f} +\sqrt{2}G_sC_{02e}+\Omega_p C_{10f},\notag 
\end{align}
where $\tilde{\omega}_{s,l}=\omega_{s,l}-i\kappa_l/2$, $\tilde{\omega}_{s,r}=\omega_{s,r}-i\kappa_r/2$, and $\tilde{\omega}_q= {\omega}_q-i\gamma_q/2$.

\begin{figure}
	\includegraphics[scale=0.26]{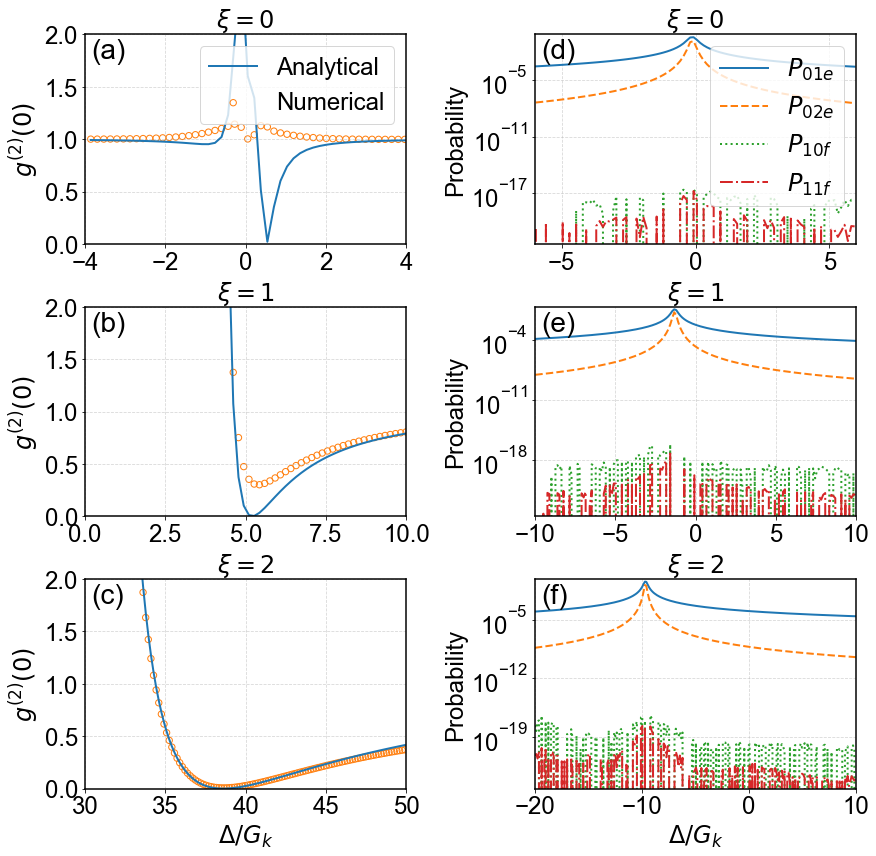}  
	\caption{The equal-time second-order correlation function $g^{(2)}(0)$ and the steady-state probabilities are plotted as functions of the normalized frequency detuning $\Delta/G_k$, with $\omega_q=10G_k$ and $\Omega_p/2\pi=0.05$~MHz. Panels (a, d) correspond to $\xi=0$, with $\omega_{s,l}=-399.805G_k$ and $\omega_{s,r}=15.016G_k$; panels (b, e) correspond to $\xi=1$, with $\omega_{s,l}=-446.250G_k$ and $\omega_{s,r}=153.887G_k$; and panels (c, f) correspond to $\xi=2$, with $\omega_{s,l}=-776.732G_k$ and $\omega_{s,r}=1142.064G_k$. Other parameters are the same as those in Fig.~\ref{figure2}. These parameter choices are made to satisfy the optimal blockade condition in Eq.~(\ref{equ35}).}\label{figure5}
\end{figure}

In a long-time limit, the system reaches its steady state, so we have $dC_{ijk}/dt=0$. From this equality, all coefficients can be directly solved. In this case,  the equal-time second-order correlation function of the magnons in the right sphere can be calculated as 
\begin{align}\label{equ34} 
	g^{(2)}(0)=&\frac{\langle m_{s,r}^{\dagger} m_{s,r}^{\dagger} m_{s,r} m_{s,r}\rangle}{\langle m_{s,r}^{\dagger}  m_{s,r} \rangle^2}\simeq\frac{2|C_{02e}|^2}{|C_{01e}|^4}.
\end{align}
Clearly, when ${2|C_{02e}|^2}/{|C_{01e}|^4}<1$, we have $g^{(2)}(0)<1$, i.e., the magnon blockade is realized. In particular, when $|C_{02e}|^2=0$, the perfect blockade [$g^{(2)}(0)=1$] is obtained. This corresponds the condition
\begin{align}\label{equ35}
	G_s^2+(\tilde{\omega}_{s,l}+\tilde{\omega}_q)(\tilde{\omega}_{s,l}+\tilde{\omega}_{s,r}+\tilde{\omega}_q)-\Omega^2_p=0.
\end{align}
Figures~\ref{figure5}(a)-\ref{figure5}(c) illustrate that perfect magnon blockade can, in principle, be achieved under the condition given by Eq.~(\ref{equ35}) (solid curves). However, numerical simulations based on the master equation reveal that, for small squeezing parameters such as $\xi=0$ (no squeezing) and $\xi=1$, perfect blockade cannot be realized. This discrepancy arises because Eq.~(\ref{equ35}) is derived under the assumption that quantum jump terms can be neglected. When the squeezing is weak, the effective enhancement of the magnon-photon coupling is insufficient to suppress quantum jump effects, and consequently, the blockade predicted by the analytical approach fails to manifest in practice.

As the squeezing parameter increases, for instance to $\xi=2$, the numerical and analytical results converge, confirming that the condition in Eq.~(\ref{equ35}) accurately predicts the regime of perfect blockade. This behavior is further corroborated by examining the occupation probabilities of the states $|0_l,1_r,e_q\rangle$, $|0_l,2_r,e_q\rangle$, $|1_l,0_r,f_q\rangle$, and $|1_l,1_r,f_q\rangle$ in Figs.~\ref{figure5}(d)-\ref{figure5}(f). It is evident that all these probabilities decrease substantially as the squeezing parameter increases, indicating that the system predominantly remains in the ground state $|0_l,0_r,e_q\rangle$ for larger $r$. This observation is fully consistent with the trends observed in Figs.~\ref{figure5}(a)-\ref{figure5}(c), indicating that perfect magnon blockade can be realized under the condition in Eq.~(\ref{equ35}) only when the squeezing parameter is sufficiently large.

\begin{figure}
	\includegraphics[scale=0.26]{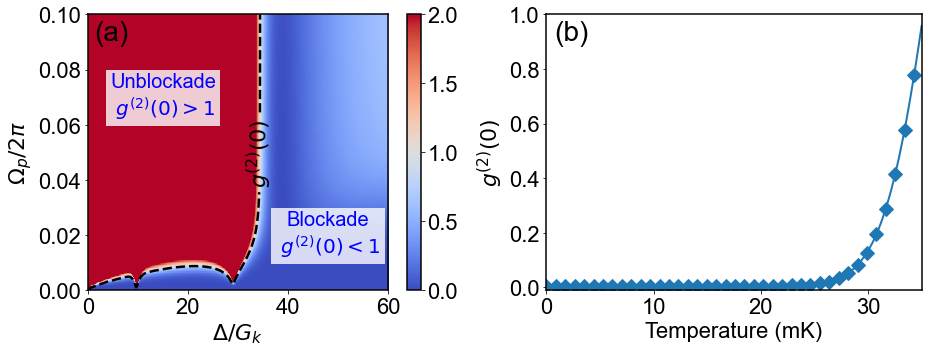}  
	\caption{(a) The equal-time second-order correlation function $g^{(2)}(0)$ as a function of  the normalized frequency detuning $\Delta/G_k$ and the amplitude $\Omega_p/2\pi$ of the probing field. (b) The equal-time second-order correlation function $g^{(2)}(0)$ as a function of the bath temperature with $\Omega_p/2\pi=0.05$ MHz. Both (a) and (b) are plotted under the condition in Eq.~(\ref{equ35}) and the squeezing parameter is $\xi=2$. Other parameters are the same as those in Figs.~\ref{figure5}(c) and~\ref{figure5}(f).}\label{figure6}
\end{figure}

We then examine the behavior of $g^{(2)}(0)$ as a function of the amplitude $\Omega_p$ of the probing field and the normalized detuning $\Delta/G_k$, as shown in Fig.~\ref{figure6}(a), under the condition given by Eq.~(\ref{equ35}). Here, the red region corresponds to the unblockade regime [$g^{(2)}(0)>1$], while the blue region corresponds to the blockade regime [$g^{(2)}(0)<1$]. The black dashed line marks the boundary $g^{(2)}(0)=1$. It is evident that for $\Delta/G_k<30$, achieving magnon blockade requires a very weak probing field with $\Omega_p/2\pi<0.01$. By contrast, for $\Delta/G_k>30$, the blockade can be sustained over a much wider range of $\Omega_p$.  In Fig.~\ref{figure6}(b), we further investigate the temperature dependence of $g^{(2)}(0)$. Remarkably, we find that nearly perfect magnon blockade remains robust against the bath temperature below $T\approx 30$~mK. However, once the temperature exceeds this threshold, $g^{(2)}(0)$ increases rapidly, and for $T>35$~mK the blockade is broken, i.e., $g^{(2)}(0)>1$.

\section{conclusion}\label{s6}

We present a scheme for realizing strong and tunable tripartite interactions in hybrid quantum magnonics by coupling a three-level NV center to driven magnons in two YIG spheres. By eliminating the ground state of the NV qutrit, an effective interaction emerges between the magnons and an NV qubit formed by the excited states. Incorporating Kerr nonlinearities and employing a squeezing transformation, we show that the tripartite interaction and cooperativity can be exponentially enhanced by joint squeezing, providing an additional $\exp(\xi)$ gain compared with conventional single-squeezing schemes. This mechanism substantially relaxes experimental requirements, accelerates system dynamics, and enables fast generation of tripartite entanglement. Furthermore, we demonstrate noise-resilient perfect magnon blockade under the enhanced interaction. Our results suggest that the NV center represents a promising interface for engineering many-body interactions in quantum magnonics, offering a versatile platform for exploring fundamental quantum phenomena such as entanglement and correlations. We emphasize that while our work focuses on NV-magnon-magnon tripartite coupling, the proposed scheme is, in principle, extendable to spin-magnon-phonon~\cite{pan2025tripartite,hei2023enhanced} and superconducting-spin-magnon systems~\cite{kounalakis2022analog}.

This work was supported by the Natural Science Foundation of Zhejiang Province (GrantNo. LY24A040004), Zhejiang Province Key R\&D Program of China (Grant No. 2025C01028),  and Shenzhen International Quantum Academy (Grant No. SIQA2024KFKT010).

\end{document}